\documentclass[preprint,showpacs,epsfig,eqsecnum,aps]{revtex4}
\usepackage{graphicx}
\usepackage{epsfig}
\begin{document}
\title{Expressions for the number of $J=0$ pairs in even--even Ti isotopes}

\author{L. Zamick$^{a)}$, A. Escuderos$^{b)}$, S.J. Lee$^{c)}$, 
A.Z. Mekjian$^{a)}$, E. Moya de Guerra$^{b)}$, A.A. Raduta$^{d)}$ 
and P. Sarriguren$^{b)}$}

\affiliation{$^{a)}$ Department of Physics and Astronomy, Rutgers University,
Piscataway, New Jersey USA 08855-8019}
\affiliation{$^{b)}$Instituto de Estructura de la Materia,
Consejo Superior de Investigaciones Cient\'{\i }ficas,
Serrano 123, E-28006 Madrid, Spain }
\affiliation{$^{c)}$Department of Physics and Institute of Natural Sciences,
Kyung Hee University, Suwon, KyungGiDo, Korea}
\affiliation{$^{d)}$Department of Theoretical Physics and Mathematics,
Bucharest University, P.O.Box MG11, Romania and Institute of Physics and 
Nuclear Engineering, Bucharest, P.O.Box MG6, Romania}

\begin{abstract}

We count the number of pairs in the single $j$-shell model of  
$^{44}$Ti for various interactions. For a state of total angular
momentum $I$, the wave function can be written as
$\Psi=\sum_{J_P\, J_N} D(J_P\, J_N) [(j^2)_{J_P}(j^2)_{J_N}]_I$,
where $D(J_P\, J_N)$ is the probability amplitude that the protons couple
to $J_P$ and the neutrons to $J_N$. For $I=0$ 
there are three states with ($I=0,\,T=0$) and one with ($I=0,\,T=2$). The
latter is the double analog of $^{44}$Ca. In that case ($T=2$), the 
magnitude of $D(JJ)$ is the same as that of a corresponding
two-particle fractional parentage coefficient. In counting the number 
of pairs with an even angular momentum $J$ we find a new relationship
obtained by diagonalizing a unitary nine-j symbol.
We are also able to get results for the `no-interaction' case for
$T=0$ states, for which it is found, e.g., that there are less ($J=1,\,T=0$)
pairs than on the average. Relative to this `no-interaction case',
we find for the most realistic interaction used that there
is an enhancement of pairs with angular momentum $J=0,2,1$ and 7, 
and a depletion for the others. Also considered are interactions in
which only the ($J=0,\,T=1$) pair state is at lower energy, only the 
($J=1,\,T=0$) pair state is lowered and where both are equally lowered,
as well as the $Q\cdot Q$ interaction.
We are also able to obtain simplified formulae for the number of $J=0$ pairs 
for the $I=0$ states in $^{46}$Ti and $^{48}$Ti by noting that the unique state
with isospin $|T_z|+2$ is orthogonal to all the states with isospin $|T_z|$.
\end{abstract}

\maketitle

\section{Introduction}
\label{sec:level1}

The plan in this work is to obtain, wherever possible, simplified
expressions for the number of pairs of particles of a given
angular momentum $J_{12}$ in the Ti isotopes in a single $j$-shell model.
We shall show that such simplified expressions can be obtained
for all even $J_{12}$ pairs in $^{44}$Ti and for all $J_{12} = 0$
pairs in $^{44}$Ti, $^{46}$Ti and $^{48}$Ti.
In a previous work \cite{moya1}, we performed calculations of the
number of pairs but did not derive simple expressions for a given $J_{12}$.
Also in this work, unlike the previous one, we compare our results
with what we call the no-interaction case which will be described later.

In the single $j$-shell model, $^{44}$Ti consists of two valence protons
and two valence neutrons in the $f_{7/2}$ shell.
The allowed states for two identical particles have angular momenta
$J=0,2,4$ and 6 and isospin $T=1$. For a neutron--proton pair, we can 
have these and also states of isospin $T=0$ with angular momenta
$J=1,3,5$ and 7. In other words, for even $J$ the isospin is one
and for odd $J$ the isospin is zero.

The wave function of a given state of total angular momentum $I$ can
be written as

\begin{equation}
\Psi=\sum_{J_P\, J_N} D^I(J_P,\, J_N) \left[ \left( j^2_\pi \right) ^{J_P}
\left( j^2_\nu \right) ^{J_N}\right] ^I\, .
\end{equation}
In the above, $D(J_P,\, J_N)$ is the probability amplitude that the protons
couple to angular momentum $J_P$ and the neutrons to $J_N$. The 
normalization condition is 
 
\begin{equation}
\sum_{J_P\, J_N} \left[ D^I(J_P,\, J_N)\right]  ^2=1,
\end{equation}
and the orthonormality condition is

\begin{equation}
\sum_{\alpha} D^\alpha(J_P,\, J_N) D^\alpha(J^\prime_P,\, J^\prime_N)
= \delta _{J_P,J^\prime_P} \delta_{J_N,J^\prime_N} ,
\label{orthon}
\end{equation}
where the sum is over the different eigenfunctions.

For states of angular momentum $I=0$, 
$J_P$ must be equal to $J_N$ ($J_P=J_N\equiv J$)

\begin{equation}
\Psi (I=0)=\sum_{J} D(JJ) \left[ \left( j^2_\pi \right) ^{J}
\left( j^2_\nu \right) ^{J}\right] ^0\, .
\end{equation}

In the single $j$-shell configuration of $^{44}$Ti, there are three $I=0$
states of isospin $T=0$ and one of isospin $T=2$. The latter is the double
analog of a state in $^{44}$Ca, i.e., of a state of four neutrons in the
$f_{7/2}$ shell. For the unique ($I=0,\,T=2$) state in $^{44}$Ti, the 
magnitudes of the $D(JJ)$'s
are the same as those of two-particle coefficients of fractional parentage

\begin{equation}
D(JJ)_{(I=0,\,T=2)}=\left[ j^2 J j^2 J |\} j^4 0\right] .
\end{equation}
We thus have for ($I=0,\,T=2$)
\begin{equation}
D(00)=0.5,\quad D(2,2)=-0.3727,\quad D(4,4)=-0.5,\quad D(6,6)=-0.600
\end{equation}

For the ($I=0,\,T=0$) states, however, the $D$'s do depend on the interaction.
We show in Table 1 the values of the $D(JJ)$'s for the lowest energy
state for the following interactions

\begin {itemize}

\item A : ($J=0,\,T=1$) pairing. All two-particle states are degenerate 
except ($J=0,\,T=1$), which is lowered relative to the others.

\item B : ($J=1,\,T=0$) pairing. Only ($J=1,\,T=0$) is lowered.

\item C : Equal $J=0$ and $J=1$ pairing. Both ($J=0,\,T=1$) and ($J=1,\,T=0$) 
are lowered by the same amount.

\item D :  $Q.Q$ interaction.

\item E : Spectrum of  $^{42}$Sc. This is the same as the
McCullen, Bayman and Zamick (MBZ) calculation \cite{mbz},
except that the correct spectrum of  $^{42}$Sc is used (some of the $T=0$
states were not known in 1964). We equate the matrix elements
\begin{equation}
\left< \left( f_{7/2}\right) ^2_J \left|V \right| \left( f_{7/2}\right) ^2_J 
\right> \, ,
\end{equation}
with $E(J)$, the excitation energy of the lowest state of angular momentum
$J$ in $^{42}$Sc. The experimental values for $J=0$ to $J=7$ are (in MeV)
0.0, 0.6111, 1.5863, 1.4904, 2.8153, 1.5101, 3.2420, and 0.6163, respectively.
Note that the three lowest states have angular momenta $J=0,1,7$.

One can add to all those numbers a constant equal to the pairing energy
$E(^{42}$Sc$)+E(^{40}$Ca$)-E(^{41}$Sc$)-E(^{41}$Ca$)$. The value is $-3.182$ MeV.
However, adding this constant will not affect the spectrum or wave functions
of $^{44}$Ti.

Note that for the even-$J$ states of $^{42}$Sc the isospin is one, while
for the odd-$J$ states the isospin is zero.

The eigenvalues and eigenfunctions of interaction E are given in Table 2. 
\end{itemize}

\section{The number of pairs in $^{44}$Ti} 
\label{secf:level2}

In this work we will use notation $A$ for the number of valence nucleons
and $n$ for the number of valence neutrons.
For Ti isotopes, $A = n + 2$.

As previously noted \cite{moya1} for a system of $A$ valence nucleons with total
isospin $T$, we have the following result for the number of pair states:
\begin {itemize}
\item Total number of pair states is $A(A-1)/2$.

\item Number with isospin $T_{12}=0$ is $A^2/8+A/4-T(T+1)/2$.

\item Number with isospin $T_{12}=1$ is $3A^2/8-3A/4+T(T+1)/2$.
\end{itemize}

Hence, for the $T=0$ state of $^{44}$Ti ($A=4$) we have three $T_{12}=0$ 
pairs and three $T_{12}=1$ pairs. For the $T=2$ state, however, we have six 
$T_{12}=1$ pairs. The important thing to note is that the number of pairs
does not depend on the two-body interaction, except for the fact that
it conserves isospin.

In $^{44}$Ti the number of pairs ($nn$, $pp$ and $np$) with total angular 
momentum $J_{12}$ ($J_{12}=0,1,2,3,4,5,6,7)$ is given by

\begin{equation}
2\left| D(J_{12}J_{12}) \right| ^2 \delta_{J_{12}, even} + 
\left| f(J_{12})\right| ^2\, ,
\end{equation}
with 
\begin{equation}
f\left( J_{12}\right) = 2 \sum_{J_P} U_{9j} (J_P,\, J_{12}) D(J_PJ_P)\, ,
\end{equation}
where we introduce the abbreviated symbol $U_{9j}$ to represent the unitary
$9j$ symbol,
which can also be written in terms of a $6j$ symbol,

\begin{eqnarray}
U_{9j}(J_PJ_{12})&=&\left< (j^2)J_P(j^2)J_P |(j^2)J_{12}(j^2)J_{12} \right> ^0 
\nonumber \\
&=&(2J_P+1)(2J_{12}+1)
\left\{ \matrix{j & j & J_P \cr j & j & J_P \cr J_{12} & J_{12} & 0} \right\}\,
\nonumber \\
 &=& (-1)^{1+J_P+J_{12}} \sqrt{(2J_P + 1)(2J_{12} + 1)} 
     \left\{\matrix{j & j & J_P \cr j & j & J_{12} }\right\} .
\end{eqnarray}
A derivation of the results up to now in this section is given in Appendix A.
Since the last publication \cite{moya1}, we have found a relationship
which in some cases simplifies the expression. The relationship
pertains only to even $J_{12}$
\begin{eqnarray}
\sum_{J_P} U_{9j}(J_PJ_{12}) D(J_PJ_P) &=& D(J_{12}J_{12}) /2 \quad {\rm for}
\quad T=0 \, ,\nonumber\\
 & = &  -D(J_{12}J_{12}) \quad {\rm for} \quad T=2\, .
\label{U9D}
\end{eqnarray}
Some useful relationships that we exploit to get this result are:
\begin{equation}
\sum_{J_{12}} U_{9j}(J_PJ_{12}) U_{9j}(J^\prime_PJ_{12}) = \delta_{J_P,
J^\prime_P}\, ,
\end{equation}

\begin{equation}
\sum_{J_{12}} U_{9j}(J_PJ_{12}) U_{9j}(J^\prime_PJ_{12}) (-1)^{J_{12}}
= -(-1)^{J_P-J^\prime_P}U_{9j}(J_PJ^\prime_P) \, .
\end{equation}

This relationship (\ref{U9D}) does not depend upon which isospin conserving
interaction is used. Using this result for even values of $J_{12}$, we find the
following:
\begin{equation}
\left| f(J_{12}) \right|^2 = \left| D(J_{12}J_{12}) \right|^2,
\end{equation}
and hence
\begin{equation}
{\rm number \;\; of}\;\; nn\;\; {\rm pairs}\;\;=\;\;
{\rm number \;\; of}\;\; pp\;\; {\rm pairs}\;\;=\;\;
{\rm number \;\; of}\;\; np\;\; {\rm pairs}\;\; =D(J_{12}J_{12})^2 .
\label{pairs-ti44}
\end{equation}

One way of looking at eq.~(\ref{U9D}) is to say that we can also write the wave
function as 
\begin{equation}
{\rm N}\sum_ {J,even} D(JJ) \left[ \left( \pi(1)\nu(2)\right) ^J \left( \pi(3)
\nu(4)\right) ^J \right]^{I=0}\, , 
\end{equation}
where N is a normalization factor.

We do not have a
corresponding simple expression for odd $J_{12}$. However, the total number of 
odd $J_{12}$ pairs must be equal to 3, the same as the total number of even 
$J_{12}$ pairs.

We can prove the relationship (2.4)  by regarding the unitary 9j symbol
as a four by four matrix where $J_P$ and $J_{12}$ assume only even values
($0,2,4,6$), despite the fact that $J_{12}$ can also assume odd values.
The eigenvalues of this matrix are $-1$ (singly degenerate) and 
$0.5$ (triply degenerate). The eigenvalue $-1$ corresponds to the 
($J=0,\, T=2$) state of $^{44}$Ti and indeed the values of $D(JJ)$ are 
identical to those obtained with a charge independent Hamiltonian and are given
in the last column of Table~1. As previously mentioned, these are the 
two-particle coefficients of fractional parentage \cite{moya1}.

The triple degeneracy with eigenvalue 0.5 corresponds to the three $T=0$ 
states being degenerate with this unitary $9j$ hamiltonian. This means that
any linear combination of the three $T=0$ states is an eigenvector. 
We can obtain the eigenvalues above without an explicit diagonalization.
This is shown in Sec.~\ref{secuni}

\subsection{Results for the (I=0, T=2) state}

Since the ($I=0,\,T=2$) state is unique, we will give the results for this case
first. Since the $^{44}$Ti $T=2$ state is the double analog of $^{44}$Ca
 \cite{RaZa},
a system of four identical particles, each pair must have even $J_{12}$.
The number of pairs is $6|[(j^2)J_{12}(j^2)J_{12}|\}j^40]|^2$, i.e., 
proportional to the square of the two-particle coefficient of fractional 
parentage. The number of pairs is
1.5 for $J_{12}=0$ and $\frac{2J_{12}+1}{6}$ for $J_{12}=2,4,6$.
This is also the result for $^{44}$Ca. Hence, even though the $I=0$ ground 
state of $^{44}$Ca has angular momentum zero and seniority zero, there are 
more $J_{12}=6$ pairs in $^{44}$Ca than there are $J_{12}=0$ pairs.
This should not be surprising. As noted by Talmi \cite{talmi} for the
simpler case of a closed neutron shell, i.e. $^{48}$Ca, the number of pairs 
with angular momentum $J$ is
equal to $2J+1$. There is only one $J=0$ pair in  $^{48}$Ca.

\subsection{Number of pairs for all states}

We can count the number of pairs for all the four $I=0$ states (three with
isospin $T=0$ and one with $T=2$).
Using the orthonormality condition (\ref{orthon}),
we eliminate the $D$'s and find
\begin{equation}
({\rm Number\ of\ pairs}) /4 = \frac{1}{2} \delta_{J_{12}, even}+\frac{1}{2}
\left[ 1-U_{9j}(J_{12}J_{12})\right] \, .
\end{equation}

\noindent The values for $T_{12}=1$ are 

0.9375 for $J_{12}=0$;  0.8542 for $J_{12}=2$; 0.9375 for  $J_{12}=4$;
and  1.0208 for $J_{12}=6$.

The total sum is 3.75.

\noindent The values for $T_{12}=0$ are 

0.3244 for $J_{12}=1$; 0.6761 for $J_{12}=3$;  0.7494 for $J_{12}=5$;
and  0.5001 for $J_{12}=7$.

The total sum is  2.25.

\subsection{Results for the T=0 ground state of $^{44}$Ti including
the no-interaction case}

In Table 3 we give results for the number of pairs for the five
interactions defined above. We also consider the `no-interaction'
case. This is obtained by getting the total number of pairs for all
three $T=0$ states and dividing by three.

It should be noted that for odd angular momentum $J_{12}$, the pair must
consist of one proton and one neutron. For even $J_{12}$, one third of 
the pairs consists of 2 protons, one third of 2 neutrons and one third of
a neutron and proton. 
We start with the `no-interaction'
result in the last column. Since there are six pairs and eight $J_{12}$'s,
if there were an equal distribution, then we could assign 0.75 pairs
to each angular momentum. This serves us as a good basis for comparison.
We find that, even in the `no-interaction' case, the results do depend
on $J_{12}$. The minimum number of pairs comes with the 
($J_{12}=1,\,T_{12}=0$)
case, only 0.432. This is of interest because there has been a lot of
discussion in recent times about ($J_{12}=1,\,T_{12}=0$) pairing. We 
start out at least with a bias against it. The maximum number of pairs
in the `no-interaction' case is for ($J_{12}=5,\,T_{12}=0$), a mode that
has largely been ignored.

However, of greater relevance is what happens to the ground state
wave function when the interaction is turned on. Therefore, we compare
the `no-interaction' case with the {\it Spectrum of} $^{42}$Sc interaction 
case E. We see striking differences. Relative
to the `no-interaction' case, there is an increase in the following number
of pairs: a) $J_{12}=0$ from 0.75 to 1.8617; b) $J_{12}=2$ from 0.861 to 
0.9458; c) $J_{12}=1$ from 0.432 to 0.6752; and d)$J_{12}=7$ from 0.667 
to 1.8945. Since the sum of all pairs in both cases is six, there must be 
a decrease in the number of pairs with the other angular momenta, and there 
is. For example, the number of  $J_{12}=6$ pairs decreases from 0.639 to 
0.0457 and the number of $J_{12}=5$ pairs from 1.00 to 0.1587. There is
also a large decrease in the number of $J_{12}=4$ and $J_{12}=3$ pairs.

The results with $Q.Q$ interaction concerning the pairs distribution with  odd
angular momenta  are qualitatively similar to the correct spectrum
of $^{42}$Sc. It is remarkable that the number of pairs with $J_{12}=0$ is almost equal to that
of pairs with $J_{12}=2$. The number of pairs of other angular momenta is almost negligible.
Thus the total number of $J=0$, 2 pairs is 2.91 out of the total of 3 even pairs.

Looking at the wave functions for $Q.Q$ (D) and the realistic interaction (E) in Table 1,
we see the dominance of $J=0$ and $J=2$ couplings for the neutrons and protons.
The percentages of the higher angular momentum couplings ($J=4$ and $J=6$) are 
only 2.9\% for $Q.Q$ and 6.4\% for
the realistic case E.
This is in accord with the Interacting Boson Model IBM2 \cite{Iache}, where
only $s$ and $d$ bosons are considered. With the simpler schematic interactions
A and B, the percentages are much higher.

We next look at the schematic interactions in the first three columns.
For the ($J=0,\,T=1$) pairing interaction, there are 
a lot of $J_{12}=0$
pairs (2.25) but very few $J_{12}=1$ pairs (0.250). The number of
$J_{12}=7$ pairs is fairly large (1.250).

For the ($J=1,\,T=0$) pairing interaction there are, as expected, a lot 
of $J_{12}=1$ pairs (1.297) and relatively few $J_{12}=0$ pairs (0.433). 
But still there is a substantial number of $J_{12}=7$ pairs (1.311). 
However, if we examine the wave function for this case it is very different 
from that of correct spectrum of $^{42}$Sc and this case represents a rather 
unrealistic ground state wave function.

For the column corresponding to equal ($J=0,\,T=1$) and ($J=1,\,T=0$) 
pairing, we get much 
better agreement in the wave function as compared with the case of correct spectrum of
$^{42}$Sc. The number of
$J_{12}=0$ pairs is 2.043 as compared with 1.862 from the case E from Table 3. For $J_{12}=1$
the values are 0.618 and 0.675, and for  $J_{12}=7$ they are 1.654 and 1.895.
Amusingly, when we lower the $J=0$ and $J=1$ matrix elements together
we get more $J_{12}=7$ pairs than we do when we lower each one separately
(1.250 and 1.311).
There is one main deficiency in the $(J=0\,+\,J=1)$ case: the number of
($J=2,\,T=0$) pairs is only 0.492 as compared with 0.9458 for the 
{\it Spectrum of $^{42}$}Sc case. The
enhancement is undoubtedly due to the quadrupole correlations in the
nucleus, an important ingredient which is sometimes forgotten when
all the emphasis is on ($J=0,\,T=1$) and ($J=1,\,T=0$) pairing. However,
if one restricts oneself to ($J=0,\,T=1$) and ($J=1,\,T=0$), then equal 
admixtures in the interaction yield much more realistic results than do
either one of them.

For the $T=2$ state of $^{44}$Ti, the double analog of $^{44}$Ca (a system
of particles of one kind), the number of $J=6$ pairs is the largest. This
suggests that the more deformed the state, (and the $T=0$ ground state of
$^{44}$Ti is certainly more deformed than the ground state of $^{44}$Ca), 
the less the number of high angular momentum pairs with the exception of 
$J$(maximum)=7. Or to put it in another way, the more spherical the state
the higher the number of high angular momentum pairs.
It is interesting that going from the $T=0$ ground state to the $T=2$ excited 
state only the $T=1$ pairs can contribute to excite the system. Therefore, the 
6 pairs which in the ground state were distributed among eight angular momenta 
($J_{12}=0,1,2,3,4,5,6,7$) are distributed in the excited state among four even
angular momenta ($J_{12}=0,2,4,6$). Note that they are not distributed 
uniformly but with  weights depending on angular momentum carried by the pair. 
Thus, the numbers of pairs in the states of angular momenta equal to 0 and 4 
are increased by a factor 2 comparing them with those corresponding to the 
non-interacting case, while for $J_{12}=6$ the  factor is 3.391. Note that the
number of pairs with angular momentum equal to 2 is decreased. It is worth 
noticing that here the favored configuration is that with the isospin ($T=2$) 
completely aligned and the angular momenta ($J=6$) completely aligned within 
the pairs, the angular momenta of pairs being antialigned.

From a previous study of `two to one' relationships  between the excitation 
energies for double and simple analog states in the shell $f_{7/2}$~
\cite{mbz,devi,RaZa}, it is noted that if we write the wave function for 
$^{43}$Ti ($^{43}$Sc) as
$$
\psi^I=\sum_{J_P} C(J_P) \left[ \left( j^2\right) ^{J_P} j_\nu \right] ^I \, ,
$$
then the $C(J_P)$'s for the $I=j$ states are identical to the $D(J_P,J_P)$'s 
for corresponding $I=0$ states in $^{44}$Ti and the
eigenvalues are one half of those. From this it follows that the number of
pairs of a given $J_{12}$ for $^{43}$Ti in a single $j$-shell calculation
is one half of those in Table~3.

\section{The number of $J_{12}=0$ pairs in $^{44}$Ti, $^{46}$Ti 
and $^{48}$Ti}
\label{seciso}

In this section we show that we can develop things further and get 
the number of $J_{12}=0$ $T_{12}=1$ pairs not only in $^{44}$Ti,
but also in $^{46}$Ti and $^{48}$Ti.

\subsection{{\em np} pairs}
\label{seciso-np}

We let $n$ be the number of valence neutrons in a given Ti isotope.
For $^{44}$Ti, $^{46}$Ti and $^{48}$Ti $n=2$, 4 and 6 respectively.
Of course the number of valence nucleons is $A = n + 2$.

The wave function for a state with total angular momentum $I=0$ is
\begin{eqnarray}
 \psi^{I=0} = \sum D(J,Jv) [(j^2)^J (j^n)^{Jv}]^{I=0}.
\end{eqnarray}
Here $v$ is the seniority quantum number.
In the Ca isotopes there is only one state for each $(Jv)$ pair.
In $^{42}$Ca we have $Jv$ pairs (00), (22), (42) and (62).
In $^{44}$Ca they are (00), (22), (42), (62), (24), (44), (54) and (84).
In $^{46}$Ca they are (00) (22) (42) and (62).

For states with total angular momentum $I=0$ in the even--even Ti isotopes,
the possible isospins in the single $j$-shell model space are
$T_{min} = |N-Z|/2$ and $T_{max} = T_{min} + 2$.
There are no $I=0$ states with $T = T_{min} + 1$.

Thus, for $^{44}$Ti we have three $I=0$ states with isospin $T=0$
and one with isospin $T=2$;
for $^{46}$Ti, five with isospin $T=1$ and one with isospin $T=3$;
for $^{48}$Ti, three with isospin $T=2$ and one with isospin $T=4$.

The formula for the number of $np$ pairs with angular momentum $J_{12}$
in a state of total angular momentum $I=0$ is
\begin{eqnarray}
 N_{np}(J_{12}) = 2n \left|\sum_J D(J, Jv) [j^{n-1}(J_0)j|\}j^nJv]
        \sqrt{(2J+1)(2J_{12}+1)}
    \left\{\matrix{j & j & J \cr j & J_0 & J_{12}} \right\}\right|^2 .
        \label{nppair}
\end{eqnarray}
In the above we have a coefficient of fractional parentage (cfp)
which is needed to separate one neutron from the others.
The six-$j$ symbol is needed, of course, to combine this neutron
with a proton in order to form an $np$ pair with angular momentum $J_{12}$.
For $^{44}$Ti the cfp is unity so the expression is simpler.

Note that for each of the even--even Ti isotopes that we are considering,
there is only one state with isospin $T_{max} = T_{min} + 2$.
Because such a state is a double analog of a corresponding state for
a system of identical particles, i.e., neutrons only in the Ca isotopes,
the amplitudes $D(J,Jv)$ of the $T_{max}$ column vector are known.
They are two-particle coefficients of fractional parentage
\begin{eqnarray}
 D^{I=0,T_{max}}(J,Jv) = [j^n J j^2 J |\} j^{n+2} 0]
\end{eqnarray}
The two-particle cfp for the Ca isotope separates a system of $(n+2)$ 
neutrons into one of two and one of $n$.
In the Ti isotopes, we separate $(n+2)$ nucleons into two protons
and $n$ neutrons.

We have also found in the past an identity which relates 
the above two-particle cfp to a one-particle cfp
\begin{eqnarray}
 [j^n (J) j |\} j^{n+1} j] = [j^n J j^2 J |\} j^{n+2} 0]
\end{eqnarray}

Now the $T_{min}$ states must be orthogonal to the $T_{max}$ states.
Thus, we get two conditions:
\begin{eqnarray}
 {\rm a) \ orthogonality} & : &
      \sum_J D^{I=0,T_{min}}(JJ) [j^n (J) j |\} j^{n+1} j] = 0  \label{ortho} \\
 {\rm b) \ normalization} & : &
      \sum_J D^{I=0,T_{max}}(JJ) [j^n (J) j |\} j^{n+1} j] = 1  \label{norma}
\end{eqnarray}
We can use these conditions to get the number of $J_{12}=0$ $T_{12}=1$ pairs
in one of the above Ti isopotes.
To do so, we find the following explicit formulae for cfp
from De Shalit and Talmi \cite{shalit} and Talmi \cite{talmi} useful
\begin{eqnarray}
 [ j^{n-1} (j v=1) j |\} j^n J=0 v=0 ] &=& 1
            \label{cfprel1}  
\end{eqnarray}
\begin{eqnarray}
 [ j^{n-1} (j v=1) j |\} j^n J v=2 ] &=& \sqrt{\frac{2(2j+1-n)}{n(2j-1)}} 
            \label{cfprel2}  
\end{eqnarray}
\begin{eqnarray}
 [ j^n (J=0 v=0) j |\} j^{n+1} j v=1 ] &=& \sqrt{\frac{(2j+1-n)}{(n+1)(2j+1)}} 
            \label{cfprel3}  
\end{eqnarray}
\begin{eqnarray}
 [ j^n (J v=2) j |\} j^{n+1} j v=1 ] &=& -\sqrt{\frac{2n(2J+1)}{(n+1)(2j+1)(2j-1)}}
            \label{cfprel4}
\end{eqnarray}
Alternatively we can use explicit expressions for two-particle cfp's given by
Lawson (his A5.57 and A5.58 equations)~\cite{l80}. They are

\begin{equation}
[j^n (J=0, v=0) j^2 0 |\} j^{n+2} 0 v=0] = \left\{ \frac{2j+1-n}{(n+1)(2j+1)}
\right\}^{1/2},
\end{equation}

\begin{equation}
[j^n (J,v=2) j^2 J |\} j^{n+2} J v=0] = - \left\{ \frac{2n(2J+1)}
{(n+1)(2j+1)(2j-1)} \right\}^{1/2} \left\{ \frac{1+(-1)^J}{2} \right\}.
\end{equation}

We define $M = \sum_{J\ge2} D(JJ) \sqrt{(2J+1)}$.
From Eqs.(\ref{ortho}), (\ref{norma}), (\ref{cfprel3}) and (\ref{cfprel4})
we find
\begin{eqnarray}
 \sqrt{\frac{(2j+1-n)}{(n+1)(2j+1)}} D(00) 
          - M \sqrt{\frac{2n}{(n+1)(2j+1)(2j-1)}}
    &=& 0  \ \ \ T=T_{min}   \nonumber \\
    &=& 1  \ \ \ T=T_{max}     \label{d00m}
\end{eqnarray}
We find $D(00) = M/3$ for $^{44}$Ti, $M/\sqrt{3}$ for $^{46}$Ti
and $M$ for $^{48}$Ti.

Now the number of pairs with angular momentum $J_{12}=0$ can
be obtained from Eq.(\ref{nppair}).
\begin{eqnarray}
 {\rm number \ of \ pairs \ } (J_{12}=0)
   = \frac{2n}{(2j+1)^2} \left|\sum_J D(JJv) [j^{n-1} (j v=1) j |\} j^n J v]
      \sqrt{(2J+1)} \right|^2
\end{eqnarray}
By using the explicit form of the cfp's in Eqs.(\ref{cfprel3}) 
and (\ref{cfprel4}), we can obtain the following relation from 
Eq.(\ref{d00m}). For $T=T_{min}$
\begin{eqnarray}
 D(00) = \frac{n}{(2j+1)} \sum_J D(JJv) (j^{n-1} (j v=1) j |\} j^n J)
             \sqrt{(2J+1)}
\end{eqnarray}
This fits in very nicely into Eq.(\ref{nppair}) to give us
one of our main results
\begin{eqnarray}
\mbox{Case 1: T=T$_{min}$} \hspace{1.5cm} 
{\rm number \ of \ {\it np} \ pairs \ } (J_{12}=0)
    = \frac{2 |D(00)|^2}{n} \hspace{3.5cm}
\end{eqnarray}
i.e., $|D(00)|^2$, $|D(00)|^2/2$ and $D(00)|^2/3$ 
for $^{44}$Ti, $^{46}$Ti and $^{48}$Ti respectively.

For $T=T_{max}$, we use the second part of Eq.(\ref{d00m}) to
obtain
\begin{eqnarray}
\mbox{Case 2: T=T$_{max}$} \hspace{0.75cm} 
{\rm number \ of \ {\it np} \ pairs \ } (J_{12}=0)
    = 2n |D(00)|^2 = \frac{2n(2j+1-n)}{(2j+1)(n+1)} .\hspace{0.5cm}
\end{eqnarray}

A similar problem of counting the pairs in a single $j$-shell was earlier 
addressed in Ref.~\cite{Engel}. Before closing this section, we would like to 
compare the results obtained in
the present paper with those in the above quoted reference.

In 1996 Engel, Langanke and Vogel \cite{Engel} worked out the number of $J_A$ 
pairs in a single $j$-shell for an isovector $(J_A=0, T=1)$ pairing 
interaction. Their main results for nuclei with ground state
with isospin $T=|T_3|=|(N-Z)/2|$ are:
\begin{eqnarray}
\langle {\cal N}_{np}\rangle &=&\frac{{\cal N}-T_3}{2T_3+3}
   \left(1-\frac{{\cal N}-T_3-3}{2\Omega}\right),
\nonumber\\
\langle {\cal N}_{pp}\rangle &=&(T_3+1)\langle {\cal N}_{np}\rangle ,\nonumber\\
\langle {\cal N}_{nn}\rangle &=&\langle {\cal N}_{pp}\rangle
    + T_3(1-\frac{{\cal N}-1}{\Omega}).
\end{eqnarray}
In the above equations, ${\cal N}$ is the
half of the total number of particles (in our notation ${\cal N}=(n+2)/2$) and
$\Omega=(2j+1)/2$. 
Their result \cite{Engel} is exact for the $J_{12}=0$, $T=1$ pairing 
interactions
and the formulae clearly show what whould happen if we dropped terms in 
$1/\Omega$.
%
 However, we obtain results for any isospin conserving interaction.
For $^{44}$Ti we obtain simple results for any even $J_{12}$ pair,
not just for $J_{12} = 0$.
For $^{46}$Ti and $^{48}$Ti, we obtain simple results for $J_{12}=0$ only.
We can show that our results, where they overlap,
agree with those of Engel {\it et al.}~\cite{Engel}.
 Indeed, as can be seen in our Table 1, the value of $D(J_{12},J_{12})$
for $J_{12}=0$ is 0.866 for the isovector pairing interation. 
This number is more precisely
$\sqrt{\frac{3}{4}}$. Thus we claim that the number of $J_{12}=0$ pairs 
with this interaction is 3/4.
This is what one gets with their formula for $\langle {\cal N}_{np}\rangle$ for
$N=Z$, $T=0$ and $\Omega=\frac{2j+1}{2}=4$.
For $T=2$ the number of even $J_A$ pairs that we get is 1.5 for $J_{12}=0$ 
and $(2J_{12}+1)/6$ for $J_{12}=2$, 4 and $6$. However, 
the above authors do not give formulae for the $T=T_3+2$ case.

\subsection{{\em nn} pairs}
\label{seciso-nn}

For the case of $^{46,48}$Ti, the number of $nn$ pairs with angular momentum 
$J_B$ can be obtained by means of the following expression:

\begin{equation}
\sum_{J_A v_A} \sum_{Jvv'} D^0(J,Jv) D^0(J,Jv') \frac{n(n-1)}{2}
[j^{n-2} J_A v_A j^2 J_B | \} j^n Jv] [j^{n-2} J_A v_A j^2 J_B | \} j^n Jv'].
\end{equation}
For $J_B=0$, we can get a simpler formula with the help of an equality that we
can find in de~Shalit and Talmi~\cite{shalit}
\begin{equation}
[j^{n-2} J v j^2 0 |\} j^n Jv]^2 = \frac{(n-v)(2j+3-n-v)}{n(n-1)(2j+1)},
\end{equation}
with $2j+1-v \ge n \ge v+2$. Thus,
\begin{equation}
\mbox{number \ of \ $nn$ pairs ($J_B=0$)} =
\frac{1}{2(2j+1)} \sum_{Jv} |D^0(J,Jv)|^2 (n-v)(2j+3-n-v).
\end{equation}
So, for $^{46}$Ti, we get
\begin{equation}
\mbox{number \ of \ $nn$ pairs} = 
\frac{1}{2} \left\{ 3|D^0(00)|^2 + \sum_{J=2,4,6} |D^0(J,Jv=2)|^2 \right\}.
\end{equation}
For $^{48}$Ti, because of the normalization of the wave function, we get
a simpler expression:

\begin{equation}
\mbox{number \ of \ $nn$ pairs} = |D^0(00)|^2 + 0.5.
\end{equation}

\noindent And for $^{44}$Ti, we already know the number of $nn$ pairs coupled to $J=0$ 
from eq.~(\ref{pairs-ti44}):

\begin{equation}
\mbox{number \ of \ $nn$ pairs }= |D^{I=0}(0,0)|^2.
\end{equation}

\subsection{{\em pp} pairs}
\label{seciso-pp}

For all three cases ($^{44,46,48}$Ti), the expression for the number of $pp$
pairs coupled to angular momentum $J_B$ is the same: $|D^{I=0}(J_B J_B)|^2$.
Thus, for $J=0$ pairs, we have 

\begin{equation}
\mbox{number \ of \ $pp$ pairs ($J=0$)} = |D^{I=0}(00)|^2.
\end{equation}

\section{Unifing the approaches of Sect.~\ref{secf:level2}
 and Sect.~\ref{seciso} and other works}
  \label{secuni}

In Sect.~\ref{secf:level2} we have in Eq.(\ref{U9D}) an eigenvalue equation for
the $D(JJ)$'s where the linear operator is a $9j$ symbol
(or actually a $6j$ symbol) whereas in Sect.~\ref{seciso} we have somewhat
similar looking equations derived from the fact that $T_{min}$
and $T_{max}$ states are orthogonal.
But here the linear operator is a coefficient of fractional parentage.

We can unifiy these two approaches by use of a recursion relation
for cfp's due to Redmond \cite{redmond}:
\begin{eqnarray}
 m [j^{m-1} (\alpha_0 J_0) j |\} j^m J] [j^{m-1} (\alpha J) j |\} j^m J]
   = \delta_{\alpha \alpha_0} \delta_{J J_0}
     + (m-1) \sqrt{(2J_0+1) (2J+1)}
          \nonumber \\  \times 
      \sum_{J_2\alpha_2} \left\{\matrix{J_2 & j & J_1 \cr J & j & J_0}\right\}
      (-1)^{J_0+J_1} [j^{m-2} (\alpha_2 J_2) j |\} j^{m-1} \alpha_0 J_0]
       [j^{m-2} (\alpha_2 J_2) j |\} j^{m-1} \alpha_1 J_1]
\end{eqnarray}
In particular, for $m=3$ we obtain the following expression
for the unitary six-$j$ symbol
\begin{eqnarray}
 \sqrt{(2J+1)(2J'+1)} \left\{\matrix{j&j&J' \cr j&j&J}\right\}
   = - \frac{\delta_{JJ'}}{2}
     + \frac{3}{2} [j^2 (J)j |\} j^3 j v=1] [j^2 (J') j |\} j^3 j v=1]
         \label{usixj}
\end{eqnarray}
From the fact that the eigenvalues obtained from the orthogonality 
of $T_{min}$ and $T_{max}$ states and the orthonormality of $T_{max}$
states as seen in Eqs.(\ref{ortho}) and (\ref{norma}) are $0$ and $1$,
we see that the eigenvalues of the unitary six-$j$ symbol are $-1/2$ and 1.
The latter results were of course shown in Sect.~\ref{secf:level2}.

It is very interesting to note that Rosensteel and Rowe \cite{rosen} also
found the need to diagonalize the above unitary six-$j$ symbol
(left-hand side of Eq.(\ref{usixj})).
They were addressing a different problem than ours.
Whereas we are here dealing with a system of both neutrons and protons, 
they were considering only particles of one kind, i.e., only neutrons.
Whereas we are addressing the number of $np$ pairs of a given angular
momentum, they were addressing the problem of the number of seniority 
conserving interactions in a single $j$-shell.
But the diagonalization of the same unitary $6j$ 
comes into play in both problems.


\section{Closing remarks}

In this work we have studied the effects of the nucleon--nucleon interaction
on the number of pairs of a given angular momentum in $^{44}$Ti. We have found
that, as expected,  the more attractive the nucleon--nucleon interaction is 
in a state with angular momentum $J$, the more pairs of that given $J$ will 
be found in $^{44}$Ti.
As a basis of comparison, we have defined the no-interaction case in which we
average over all three $T=0$ states (in the single $j$-shell approximation) 
in $^{44}$Ti. Even in this case, the number of $J$ pairs is not independent of 
$J$ and there
are, for example, less $J=1$ pairs than the average (0.432 vs. 6/8=0.75).
When the realistic interaction is turned on, one gets, relative to this 
no-interaction case, an increase in the number of $J=0,1,2$ and 7 pairs and
a decrease in the others. This is in accord with the fact that in $^{42}$Sc
the states with angular momentum $J=0,1,2$ and 7 are lower than the others.

The number of pairs obviously is of relevance to two-nucleon transfer experiments
and we plan to address this more explicitly in the near future. For example,
the pickup of an $np$ pair in $^{44}$Ti to the $J=1$ state in $^{42}$Sc
will be enhanced, relative to the no-interaction case by a factor of
$(0.675/0.432)^2$. Although $^{44}$Ti is radioactive, such experiments,
as well as transfer
reactions, are feasible. Indeed, there are approved proposals to perform the reaction
$^{3}$He($^{44}$Ti, p)$^{46}$V by a Berkeley group.


In the course of this work, we have found a relationship between the wave 
function coefficients $D(J,Jv)$ for the even--even Ti isotopes by exploiting 
the fact that, in the single $j$-shell, the unique state with isospin $T_{max}$
is orthogonal to all the states with isospin $T_{min}$ [Eqs.~(\ref{ortho}) and
(\ref{norma})]. This enabled us to greatly simplify the expression for the 
number of $np$ pairs with angular momentum $J_{12}=0$ in the even--even Ti 
isotopes.

In the near future, we intend to extend this analysis for a single shell with 
large angular momentum and an arbitrary number $Z=N$ of protons and neutrons. 
This would allow  us to study the competition between the $T=1$ and $T=0$ 
pairing interactions in the structure of the ground state in the medium mass 
nuclei.

The competition betwen $T=0$ and $T=1$ pairs in the ground state might be best studied with
an NT projected generalized BCS function, derived by two of us (AAR and EMG) in
Ref.~\cite{Rad}.
This project is under way and we hope to have then a more fair comparison between
various approaches.
However, based on the present results we may say that, while in the ground state the
number of the $T=1$ and $T=0$ pairs are equal to each other, in the $T=2$ state the number of the
$T=1$ pairs prevails over that of the $T=0$  pairs, which is equal to zero. 
Thus, in the present case
the change in the wave function from a mixed state of $T=0$ and
$T=1$ pairs to a pure state of $T=1$ pairs is achieved by exciting 
the system from a $T=0$ ground state to a $T=2$ state.

\section{Appendix A. The number of pairs of a given angular momentum in the
single $j$-shell in $^{44}$Ti }
\renewcommand{\theequation}{A.\arabic{equation}}
\setcounter{equation}{0}

In the single $j$-shell, there are 8 two-body interaction matrix elements
\begin{equation}
E(J) = \left< \left( f_{7/2}\right) ^2_J \left| V \right| 
\left( f_{7/2}\right) ^2_J \right> \, ,
\end{equation}
$J=0,1,...,7$. For even $J$, the isospin is $T=1$; for odd $J$, $T=0$. The 
energy of a $^{44}$Ti state can be written as $<\psi H \psi>$. This can be 
written as a linear combination of the eight two-body matrix elements $E(J)$
\begin{equation}
E(^{44}Ti) = \sum_{J=0}^7 C_J E(J)\, .
\end{equation}
We can identify $C_J$ as the number of pairs in $^{44}$Ti with a given angular 
momentum $J$
\begin{equation}
<\psi H\psi> = \sum D(J'_PJ'_N)D(J_PJ_N) \left< \left[ J'_PJ'_N\right] ^I H 
\left[ J_PJ_N\right] ^I\right> \, ,
\end{equation}

\begin{eqnarray}
&&\left< \left[ J'_PJ'_N\right] ^I H \left[ J_PJ_N\right] ^I\right> =
\left[ E\left( J_P\right) + E\left( J_N\right) \right] \delta_{J_PJ'_P}
\delta_{J_NJ'_N} \nonumber  \\
&&\qquad +4\sum_{J_AJ_B} \left< \left( j^2\right) J'_P  
\left( j^2\right) J'_N |  \left( j^2\right) J_A  \left( j^2\right) J_B
\right> ^I  \nonumber \\
&&\qquad \times \left< \left( j^2\right) J_P  
\left( j^2\right) J_N |  \left( j^2\right) J_A  \left( j^2\right) J_B
\right> ^I E(J_B) \, .
\end{eqnarray}
In the above, the first two terms are the $pp$ and $nn$ interactions and the 
last one is the $np$ interaction. The factor of 4 is due to the fact that 
there are 
4 $np$ pairs. The unitary 9j symbol recombines a proton and a neutron.
Note that $J_P$ and $J_N$ are even but $J_A$ and $J_B$ can be even or odd.

By identifying the coefficient of $E(J_B)$ as the number of pairs with
angular momentum $J_B$, we get the expression for $I=0$ (for which $J_P=J_N$)

\begin{eqnarray}
&&C_{J_B} = {\rm Number\; of}\; J_B\; {\rm pairs} = 2\left[ D(J_BJ_B)
\right] ^2 \delta_{J_B\; ,\, even}
\nonumber \\
&&\qquad +4\sum_{J_PJ_N}  D(J_PJ_N) \left< \left( j^2\right) J_P  
\left( j^2\right) J_N |  \left( j^2\right) J_B \left( j^2\right) J_B
\right> ^0 \nonumber \\
&&\qquad \times \sum_{J'_PJ'_N} D(J'_PJ'_N)
 \left< \left( j^2\right) J'_P  
\left( j^2\right) J'_N |  \left( j^2\right) J_B  \left( j^2\right) J_B
\right> ^0 \, .
\end{eqnarray}

We can rewrite this as 
\begin{equation}
{\rm Number\; of}\; J_B\; {\rm pairs} = 2\left[ D(J_BJ_B)\right] ^2 
\delta_{J_B\; ,\, even}\;
+\left| f(J_B)\right| ^2 \, ,
\end{equation}
with
\begin{equation}
f(J_B)= 2  \sum_{J_PJ_N} D(J_PJ_N) \left< \left( j^2\right) J_P  
\left( j^2\right) J_N |  \left( j^2\right) J_B  \left( j^2\right) J_B
\right> ^0 \, ,
\end{equation}
and
\begin{eqnarray}
&&\left< \left( j_1j_2\right) J_P  
\left( j_3j_4\right) J_N |  \left( j_1j_3\right) J_A  \left( j_2j_4\right) 
J_B \right> ^I \nonumber \\
&&=\sqrt{(2J_P+1)(2J_N+1)(2J_A+1)(2J_B+1)}
\left\{ \matrix{j_1 & j_2 & J_P \cr j_3 & j_4 & J_N \cr J_A & J_B & I} 
\right\} \, .
\end{eqnarray}

We can get the number of pairs for a given total isospin $T$ by using a 
simple interaction $a+b t(1)\cdot t(2)$, where $a$ and $b$ are constants. 
The value of this interaction for two particles is
\begin{equation}
a-3b/4 \; ({\rm for} \; T_{12}=0) \quad {\rm and}\quad
a+b/4 \; ({\rm for} \; T_{12}=1) \, .
\end{equation}
For $A$ valence nucleons we have ($A = n+2$ for the Ti isotopes)
\begin{eqnarray}
\sum_{i<j} \left( a+b t(i)\cdot t(j)\right) &=&
\frac{a}{2}A(A-1) +\frac{b}{2}\sum_{i,j}t(i)\cdot t(j)
-\frac{b}{2}\sum_{i} t(i)^2 \nonumber \\
&=&\frac{a}{2} A (A-1)+\frac{b}{2}T(T+1) -\frac{3}{8} A b \, .
\end{eqnarray}
We can write this as 
\begin{equation}
A_0 (a-3b/4)+A_1 (a+b/4)\, ,
\end{equation}
and identify $A_{T_{12}}$ as the number of pairs with isospin $T_{12}$. 
We then get the result of Sect.~\ref{secf:level2}:
\begin{eqnarray}
A_0 &=& \frac{A^2}{8}+\frac{A}{4}-\frac{T(T+1)}{2} \, ,\\
A_1 &=& \frac{3 A^2}{8}-\frac{3 A}{4}+\frac{T(T+1)}{2}\, .
\end{eqnarray}

\begin{center}
{\Large \bf Acknowledgments} 
\end{center}

This work was partially supported by MCyT (Spain) under contract number
BFM2002-03562, by  a U.S. Dept. of Energy Grant No. DE-FG0104ER04-02,
by Grant No. R05-2001-000-00097-0 from the Basic Research Program
of the Korea Science and Engineering Foundation,
by MEC (Romania) under the grant CERES2/24, 
and by a NATO Linkage Grant PST 978158. One of us (A.E.) is supported by a 
grant financed by the Secretar\'{\i}a de Estado de Educaci\'on y Universidades
(Spain) and cofinanced by the European Social Fund.

\vfill\eject

\vfill \eject

\begin{table}[ht]

{\bf Table 1.} Wave functions of $^{44}$Ti for various interactions:
A ($J=0,\,T=1$) pairing;  B ($J=1,\,T=0$) pairing;  
C equal $J=0,J=1$ pairing; D Q.Q interaction;  E spectrum of $^{42}$Sc.
\vskip .5cm

\begin{center}
\begin{tabular}{|c|ccccc|c|}
\cline{2-7}
\multicolumn{1}{c}{} & \multicolumn{5}{|c} {Ground state $T=0$}& $T=2$ \\
\hline
 $D(JJ)$ &$\hskip0.5cm$ A$\hskip0.5cm$ &$\hskip0.5cm$ B$\hskip0.5cm$
 &$\hskip0.5cm$ C$\hskip0.5cm$ &$\hskip0.5cm$ D$\hskip0.5cm$ &$\hskip0.5cm$ E$\hskip0.5cm$ &
 any interaction  \\
\hline
$J=0$  & 0.866 &  0.380 & 0.826 & 0.7069 & 0.7878 & 0.5   \\
$J=2$  & 0.213 &  0.688 & 0.405 & 0.6863 & 0.5617 & -0.3727 \\
$J=4$  & 0.289 &  0.416 & 0.373 & 0.1694 & 0.2208 & -0.5    \\
$J=6$  & 0.347 & -0.457 & 0.126 & 0.0216 & 0.1234 & -0.6009 \\
\hline
\end{tabular}
\end{center}
\end{table}

\vskip 1cm
\begin{table}[h]

{\bf Table 2.} 
Excitation energies [MeV] and eigenvectors of the {\it Spectrum of $^{42}$Sc} 
interaction.

\vskip .5cm

\begin{center}
\begin{tabular}{|c|c|cccc|}
\hline
excitation energies&&  $\hskip0.5cm$   0.0000  $\hskip0.5cm$  &
 $\hskip0.5cm$5.4861 $\hskip0.5cm$   & $\hskip0.5cm$ 8.2840 $\hskip0.5cm$  &
  $\hskip0.5cm$ 8.7875 $\hskip0.5cm$   \\
\hline
 &D(00)              & 0.78776 &-0.35240 &-0.50000 & -0.07248 \\
eigenvectors &D(22)              & 0.56165 & 0.73700 &0.37268  & -0.04988 \\
 &D(44)              & 0.22082 &-0.37028 & 0.50000 & 0.75109 \\
 &D(66)              & 0.12341 &-0.44219 & 0.60093 &-0.65432 \\
\hline
\end{tabular}
\end{center}
\end{table}

\vskip1cm
\begin{table}[h]

{\bf Table 3.} 
Number of pairs for the $T=0$ state of $^{44}$Ti with various interactions:
A ($J=0,\,T=1$) pairing;  B ($J=1,\,T=0$) pairing;  
C equal $J=0,J=1$ pairing; D  Q.Q interaction; E spectrum of $^{42}$Sc;
F no interaction. In G we give the number of pairs for the $T=2$ state.

\vskip .5cm

\begin{center}
\begin{tabular}{|c|cccccc|c|}
\hline
$\hskip0.5cm$ &$\hskip0.5cm$ A$\hskip0.5cm$ &$\hskip0.5cm$ B$\hskip0.5cm$ &$\hskip0.5cm$
C$\hskip0.5cm$ &$\hskip0.5cm$ D$\hskip0.5cm$ &$\hskip0.5cm$ E$\hskip0.5cm$
&$\hskip0.5cm$ F$\hskip0.5cm$ & $\hskip0.5cm$ G$\hskip0.5cm$ \\
\hline
$J_{12}=0$ & 2.250 & 0.433 & 2.045 & 1.499 & 1.862 & 0.750 & 1.500 \\
$J_{12}=2$ & 0.139 & 1.420 & 0.492 & 1.413 & 0.946 & 0.861 & 0.833 \\
$J_{12}=4$ & 0.250 & 0.320 & 0.416 & 0.086 & 0.146 & 0.750 & 1.500 \\
$J_{12}=6$ & 0.361 & 0.626 & 0.048 & 0.001 & 0.046 & 0.639 & 2.167 \\
\hline
$J_{12}=1$ & 0.250 & 1.297 & 0.618 & 0.834 & 0.675 & 0.432 & \\
$J_{12}=3$ & 0.583 & 0.388 & 0.165 & 0.156 & 0.271 & 0.902 & \\
$J_{12}=5$ & 0.916 & 0.003 & 0.564 & 0.013 & 0.159 & 1.000 & \\
$J_{12}=7$ & 1.250 & 1.311 & 1.654 & 1.996 & 1.895 & 0.667 & \\
\hline
\end{tabular}
\end{center}
\end{table}

\end{document}